\documentclass{article}[11pt]
 
\newcommand {\bea}{\begin{eqnarray}}
\newcommand {\eea}{\end{eqnarray}}
\newcommand {\be}{\begin{equation}}
\newcommand {\ee}{\end{equation}}
\newcommand {\bc}{\begin{center}}
\newcommand {\ec}{\end{center}}

\def\lsim{\mathrel{\rlap{\lower4pt\hbox{$\sim$}}
    \raise1pt\hbox{$<$}}}               
\def\gsim{\mathrel{\rlap{\lower4pt\hbox{$\sim$}}
    \raise1pt\hbox{$>$}}} 

\usepackage{amsmath, amssymb, amsfonts, amsthm}
\usepackage{stmaryrd}
\usepackage{graphicx}
\usepackage[all]{xy}
\usepackage{color}
\usepackage{bm}
\usepackage{subfigure}
\usepackage{graphicx}
\usepackage{mathabx}
\usepackage{multirow}
\usepackage{setspace}
\usepackage{epstopdf}
\usepackage{makeidx}
\usepackage{multicol}  
\usepackage{braket} 

\usepackage{geometry}
\graphicspath{{./Pictures/}}

\author{Clifford E Chafin\\\ \small{Department of Physics, North Carolina State University, Raleigh, NC 27695} \thanks{cechafin@ncsu.edu}}
\title{The Quantum State of Classical Matter I: Solids and Measurements}

\begin{document}
\begingroup
\let\clearpage\relax
\maketitle
\begin{abstract}
Using the kinematic constraints of classical bodies we construct the allowable wavefunctions corresponding to classical solids.  These are shown to be long lived metastable states that are qualitatively far from eigenstates of the true Hamiltonian.  Extensions of this give an explicit description of phonon oscillations in terms of the wavefunction itself and some consequences for the general validity of the quasiparticle picture are presented.  An intrinsic theory of quantum measurement naturally arises based on Schr\"{o}dinger evolution that is local, consistent with relativity and extends to the case of noninertial and deformable measurement devices that can have time changing internal properties.  This theory agrees with the Born interpretation in the limit of static measuring devices.  Care is given to the transport of conserved quantities during measurement.  
\end{abstract}

\section{Introduction}

The creation of condensed Bose gases has led to close to twenty years of intensive research  probing their static and dynamic behavior.  The creation of Fermionic gases and the ability to fully control their interaction strength through Fano-Feschbach resonances combined with their relatively long stability has allowed us to study ever broadening regimes of behavior.  This has opened new directions and a search for new expressions and tests of theory in the areas of superfluidity, superconductivity, other exotic quantum phases, quantum ``simulators'' of condensed matter systems, and the transition from BEC to BCS states \cite{Bloch:2007}\cite{Giorgini:2008}.  The unitary limit, where s-wave scattering lengths diverge, has attracted considerable attention as it seems to lead to simplification of many particle motions into hydrodynamic motions that seem to persist from the coldest to the the hottest systems mankind has at his disposal.  

Theoretic work on these systems was pioneered by Bogoliubov, Yang, Lee, Huang \cite{LHY:57}\cite{Bogo:57} and many others.  A natural question is how the size and dynamics of these traps are affected by temperature, pressure or other thermodynamic functions.  Implicit in such variables is that such systems undergo equilibration and, thus, independent of the history and preparation of these system.  A prominent feature of these strongly interacting Fermi systems is ``elliptic flow,'' the free expansion of a cloud that  seems strongly hydrodynamic.  In hydrodynamics, the ways vorticity can enter a system is  constrained.  In the absence of density gradients, viscosity can only pull vorticity in through the surface \cite{Batchelor}.  Quantum systems (i.e. wavefunctions) also are very restricted with respect to vorticity.  However, here vorticity for an N-body $\psi$ can only exist in the form of singular 3N-2 dimensional hypercurves where the wavefunction's norm vanishes.  For certain strongly interacting cases, like superfluid Helium, these exhibit a coherence so that these give zeros in the one body density function.  It is interesting to note that a long standing problem in hydrodynamics is the immortality of solutions, specifically the nonsingular existence of solutions to Navier-Stokes for all time \cite{Majda}.  This has been shown to reduce to the equivalent property of never forming singular vorticity in finite time.  Ironically, quantum systems only show singular vorticity for all time.  Some bridge that represents classical fluids as limits of special quantum systems might shed light on this old problem.  

Much of the work on ultracold gases has centered on finding equations of state and transport functions for hydrodynamic and thermodynamic expressions in their description\cite{Chevy:2009}.  There is, however, an alternate and under discussed point of view that such systems might be a powerful lever in the investigation of some of the old and languishing problems at the heart of physics: measurement theory and the unification of classical and quantum dynamics, ensembles as a foundation for statistical mechanics and thermodynamics, and the justification and consistency of quantum statistical mechanics.  These points are not purely academic.  To some degree, quantum mechanics, classical mechanics, and thermodynamics exist as independent pillars of a physics that are still unreconciled.  This leaves questions about their consistency in doubt and, since they are related by ansatz and prescriptions, how to proceed in nonequilibrium cases and on scales that are between classical and quantum unclear.  Ultimately, microscopic \textit{dynamics} and the theory's founding constraints should be the origin of all collective and limiting behaviors.  

To discuss the thermal and hydro properties of condensed quantum systems, it would be very useful to have a consistent quantum treatment of classical matter.  This would suggest when parallels will be justified and not.  Hydrodynamics and elasticity are traditionally classical fields though there has been an encroachment on them by quantum means to get transport and mechanical properties just as there has been a supplanting of classical statistical mechanics by quantum methods.  Although some of these calculations have been very successful, others, with a priori just as much reason to succeed, have not and the foundations of quantum statistical mechanics is far more ad hoc and conceptually vague than its classical counterpart. Indeed, many of the exciting advances in statistical mechanics in recent years \cite{Cohen:1980}\cite{Jarzynski} have been limited to classical cases with no path to finding a quantum version.  

The most fundamental of these questions is how to describe a measurement device, typically a collection of solids arranged to give discrete measurements.  If we can give a consistent and explicit quantum treatment of a solid body that generates an explanation of quantum measurement and thermodynamics it will be a promising candidate to examine these other more challenging questions.  Unfortunately, the topic of measurement theory now provokes a great aversion on the part of many physicists.  After years of philosophically heavy discussions and posturing that seemed to produce no new predictions many are, if not true positivists, more inclined to work with existing formalism than risk getting drawn down an unproductive avenue or engage in arguments that seem more about words than physics.  

The quantum theory of measurement is typically expressed via the Born interpretation, a measurement ansatz where a ``measurement device'' is associated with a linear self-adjoint operator that projects out a spectrum of possible measurements and their probabilities in a measurement ``event'' \cite{Dirac}\cite{Neumann}\cite{CoTan}.  Despite having no fundamental way to explain why certain devices are measurement devices and what their spectrum of action should be, this has been the most economical and enduring approach to the task of relating wavefunctions to measurements.  

Frustration with the current theory of quantum measurement goes back to Schr\"{o}dinger himself who said that if he knew what they were going to do with his wavefunction, he never would have written it down \cite{Jammer}.  When quantum and classical worlds seemed so vastly separated by scales, a positivist attitude seemed justified.  Strangely, it was vehemently argued that nothing more was to be asked.  This notion, embodied in the principle of complementarity, was a kind of life-goal for Bohr and became broadly accepted.  Of course, this separation of scales was destined to be temporary and, more than any argument about determinism, ontology, epistimology\ldots , eventually we will need a unified description of classical matter in the same language of the microscopic.  As an example, the action of a given measurement device should be determined by its microscopic nature.  We should not have to require classical + quantum descriptions + a ``by-fiat'' assignment of the quantum measurement action of a given classical object.  Ultimately we would like to replace ``interpretation'' with direct implication.  

In the first few sections
we take the point of view that all matter (of fixed particle number) is a wavefunction and well governed by the Schr\"{o}dinger equation.  By applying the basic constraints of classical rigid bodies to a wavefunction we derive quantum description for a solid body that includes a set of many body excitations clearly identifiable with phonons thus allowing a direct wavefunction description of, while not phonons themselves, the solid in an excited state corresponding to phonon occupancy.  This represents a departure from the usual second quantized description.  

As a by-product we find that solid bodies (and therefore much of what we call classical matter) are actually somewhat far from true eigenstates of the Hamiltonian describing them even if they are cooled to absolute zero.  They do however have the property of very long lifetimes (according to Schr\"{o}dinger dynamics).  These are the special subset of wavefunctions that exhibit the usual kinematic freedom of classical bodies as 3D objects with Newtonian dynamics.  Combining these with smaller bodies formed by evaporation and delocalization will lead to a Schr\"{o}dinger equation mediated measurement theory that is, in a proper limit, identical to the Born ansatz.  More general configurations lead to a more general theory of measurement that holds for moving (noninertial) and deforming measurement devices that can be changing in chemical/crystallographic nature, measurement efficiency and even relativistic.  

A consequence of this description we demonstrate that a ``many worldsy'' sort of measurement theory follows immediately from their interaction with delocalized particles.  The large masses of these macroscopic objects creates an effective long lasting partition of the space that can be indexed by the incident wavefunction.  Unlike approaches where the wavefunction ``collapses'' or undergoes some other rapid qualitative change (inducing locality problems, concerns about conserved quantities and back reactions on the measurement device), in this case the macroscopic world is ``sliced'' into temporal foliations by their interaction with delocalized bodies.  The many body wavefunction continues to exist but these indexed slices cease to interact on human time scales thus leading to a kind branching of paths for the observers.  Interestingly, this gives a completely local description of measurement that also allows for noninertial and otherwise dynamic measurement devices, situations that a complete theory should address but, for which, no obvious linear self-adjoint operator (LSAO) exists \cite{CoTan}.  Indeed, it is a local and ``complete'' theory of measurement in that it gives deterministic results for the location and time of events regardless of what we do to our measurement device.  It reduces quantum mechanics to a classical field theory with all probabalistic and time asymmetric features arising for observers due to the special (and nonpermanent) qualities of wavefunctions that can describe classical objects.  

In practice quantum theory allows a great degree of freedom to generate calculations.  ``Quantization'' of classical motion using effective lagrangians is often successful but not clearly consistent or derivative from more microscopic descriptions.  Second quantization provides an equivalent description of (fixed N) N-body quantum mechanics but, the most common examples involve ``quasiparticles'' when such a connection and the the nature of the coupling is not so clearly equivalent.  Momentum conserving approaches for the couplings almost always involve pseudomomentum so further justification of these coupling rules should be presented.  Field theoretic methods have been very powerful but provide less geometric intuition and more opportunities for hidden confusion and accidents to enter than almost any other method.  Without greater specificity and controls on the series resummations we will never be able to be confident in its results without confirming experimental data or Monte Carlo results.  Even then, we have no a priori restrictions on the bouquet of currently allowable calculations thus little certainty that they relate to the quantum results that follow from a description at the smallest of scales.  
We will use our specific wavefunction for solids to suggest some bounds on when the quasiparticle approach is justified and glean some intuition for how far such calculations can be pushed.  

The nature of this discussion will be occasionally qualitative and leave room for significant elaboration.  Given the great range of work covered by quantum theory, the longstanding conceptual holes in it, and the great liberty the subject has allowed in the justification of various results, this seems a justifiable necessity for such a new approach.  The corresponding ``intrinsic'' (i.e.\ following directly from the Schr\"{o}dinger equation) approaches to the subjects of thermodynamics of matter, hydrodynamics of gases and liquids, and particle number changing systems will not be covered here as they introduce further significant complications but are to be presented in  following papers.  

\section{Condensed Matter and the Classical Limit}

\subsection{Molecules and Bonds}Before we consider the case of the wavefunction of a large classical body, we consider the few body case.  The two body case is one of the few exactly solvable examples.  Separation into relative and center of mass (CM) coordinates $(x_{1},x_{2})\rightarrow(x_{rel},X_{cm})$ gives a free translating part with the net mass $M=m_{1}+m_{2}$ and internal motion given by radial oscillations and angular motion of a single particle with reduced mass $\mu=\frac{1}{m_{1}^{-1}+m_{2}^{-1}}$.  

The nature of the two body potential and particle symmetries enter here by determining the nodal structure of the ground state wavefunction at the origin of the relative coordinates.  For the case of a hydrogen atom, the function has no node here and we find that the ground state is unique and angular excitations require the presence of radial ones.  This is encompassed in the rule for the angular and magnetic quantum numbers $l<n$ and $|m|\le l$.  If the potential is for a pair of fermions or comes from composite bodies with an effective hard core repulsion then there can be a node in the ground state of $\psi_{rel}$ and angular excitations may be much less expensive than radial ones.  This explains why the specific heats of diatomic molecules occupy the angular excitations before the vibrational (or equivalently, radial) modes make a contribution whereas, for the electronic states of a Hydrogenic atom, this is not the case.  

Often the consideration of angular and vibrational modes of diatomic molecules, as that of the phonon modes of solids, is done by a quasi-classical consideration with ``quantization'' of the classical modes done as a final step.  This breaks the continuity of the argument and leaves room for doubt as to the consistency of such an analysis from a direct consideration of the total wavefunction of the system.  Many students have wondered when to insert such a magical/inspired step and if they really understood the solutions to the Schr\"{o}dinger equation after all.  

In the spirit of interpreting all the dynamics of a system in terms of its wavefunction, let us approach this differently.  The wavefunction of H$_{2}$ decomposes into a 3-D center of mass part and a 9-D internal component $\Psi=\psi_{cm}\psi_{int}$.  Including the $2^{4}$ component spin basis this gives a set of 16 12-D functions to keep track of.  
Because the electrons have such small mass compare to the protons we can, in some sense, ignore them with small correction in deriving the energy levels.  However, if the center of mass is delocalized to a much greater extent than the radius of the atom, we can have a very large relative discrepancy in the \textit{value} of $\Psi$ at any given point in $\mathbb{R}^{12}$ and the relative size of such an approximation gets worse at larger values of the particle coordinates even though the energy per density changes are small.  It turns out that a careful accounting of the center of mass coordinates and large coordinate effects arising from particular small energy changes will be central to later arguments so we will be, prima facie, pedantic about this.  

In molecular orbital theory (MO) we often consider the relative coordinate wavefunctions to be those of the electrons.  Since the reduce mass corrections are small, this is often accurate enough for such calculations.  However, when considering the antisymmetry of the electronic part of the wavefunction this does introduce some error.  
Ultimately, the antisymmetry must be done over the electrons coordinates not the weighted mix of electron and core coordinates that make up the relative coordinate wavefunctions.  

For the present case, let us place our H$_{2}$ molecule in a box of length $L$ where $d\ll L\ll \frac{m_{p}}{m_{p}-\mu_{H}}d$ to limit the CM delocalization so that MO theory approximations to the values of $\Psi$ are small.  The value $d$ here is a parameter on the order of the atomic size.  An s-orbital approximation gives, 
\begin{align}
\Psi=\hat{\mathcal{S}}_{e}\hat{\mathcal{S}}_{p}\phi_{s}(x_{1}-(X_{1}-X_{2})/2)\phi_{s}(x_{2}+(X_{1}-X_{2})/2)\psi_{rel}(X_{1}-X_{2})\psi_{cm}(X_{1}+X_{2})
\end{align}
 where the lowercase variables are the electron coordinate labels and the uppercase are the protons.  Since we assume opposite spins for both these are symmetrized by the operators $\hat{\mathcal{S}}_{e,p}$.  $\psi_{rel}(y)$ gives a strongly peaked function at $|y|=d$, the proton separation distance.  It is a spherically symmetric function that damps strongly at zero.  The electronic contribution looks like a pair of overlapping orbitals that we see in books on molecular orbital theory or band structure.  The details of the hidden coordinates are generally obscured by these pictures and models.  They inherit the rotational symmetry from $\psi_{rel}$ as we advance through the $y=(X_{1}-X_{2})$ coordinate subspace.  It is a slight inconsistency that we use the reduced mass and relative coordinates to specify the atomic orbitals but use the entire mass of the the atoms to form the molecular wavefunction.  The overlap region shifts the mass density of the electrons in a way not considered here and the the relative coordinates in the orbitals are really only partly electronic and contain small components of proton coordinate contributions. 

This begs the question of how do conservation laws relate to  symmetries in the case of many body wavefunctions.  We generally look at the classical Lagrangian and apply Noether's theorem then assume (or verify independently) the system inherits these symmetries after ``second quantization.''  It would be more satisfying to simply have a Lagrangian for the many body wavefunction to start with and derive these directly.  We will leave this (and its surprising complications) for a later section.  For now, we notice that for the interacting case and absent the large box we placed around the system, there is free translation along the center of mass coordinate.  Rotational symmetry is expressed in the rotation in the proton relative coordinate variable $y=(X_{1}-X_{2})$ and the relative wavefunction $\psi_{rel}$ gives a narrow width spherical shell.  The angular modes are thus easy to excite and their narrow radial spread justifies calculations using the classical moment of inertia in estimating their energy.  In the case of allowing a finite mass contribution of the electrons, the rotational freedom would involve simultaneous rotation of different sizes in each of the four coordinates such that the center of mass remains unaffected.  

Not making details of the atomic and molecular orbitals explicit in terms of the underlying particle coordinates clear invites a number of paradoxes such as the effects of bulk rotation on a metal or superconductor.  Consider the case of a rotating metal or superconductor.  According to band theory, as we rotate the solid, the electron band occupancy should become excited to cancel any net current of the cores.  This however requires the existence of vortices in the electron cloud.  In the case of superconductors, this is particularly expensive and difficult to cancel the field in the interior without many of them to cancel the fields of the previously introduced ones.  To resolve this, consider the case of a lone atom being moved at quantized angular velocity over a circular loop of radius $R$.  If the proton is considered to to this motion and the electron moving in a cloud about it then we expect the electron cloud will exhibit a precessing deformation from small contributions of excited angular momentum states.  The associated vorticity from these will generally be far into the low amplitude tail of the electronic wavefunction.  This small correction to the atomic orbital can then correct the underlying atomic orbitals the make up the band structure leaving no need to make expensive and complicated changes in the band structure (molecular orbital) occupancy.  

\subsection{Solids}\label{solids}

As we combine larger numbers of atoms to build solids we are left with the following questions.  Do larger numbers of particles automatically give classical bodies?  Do they necessarily lead to thermodynamically equilibrating states?  Here we will investigate the first question and leave the second for later.  MO theory and the Bloch model of solids generally involves placing atoms at specific locations (of their center of masses) as delta functions and using energy minimization over the possible atomic orbital combinations to derive the molecular orbitals or band structure.  The role of metastability of these various combinations is rarely discussed especially in terms of how these states relates to the \textit{true} eigenstates of the given nuclei and electrons.\footnote{Of course, the states of nuclei are also generally metastable and the role of variable particle number spaces in defining a basis of stationary states is not immediately obvious so we are really just investigating the first layer of such considerations.}  

On of the more lightly thrown about phrases in introductory quantum mechanics is that ``identical particles are indistinguishable.''  This is often used in conjunction with semi-classical pictures of atoms at specific locations.  STM images of a solid surface which show atoms as well defined peaks might seem to contradict this.  These atoms can have the same nuclei and atomic structure as their neighbors yet are ``distinguished'' by their different locations.  Indeed the atoms at a surface are different in their electronic structure from those in the bulk of the solid and there are gradations of variations as we move away from the surface and variations introduced by time dependent thermal fluctuations.  The essential message of ``indistinguishability'' is that the many body wavefunction needs to have specific symmetries imposed on it.  Such vague language confuses the particle coordinate label and the particle location.  This probably arose from the long and confusing arguments regarding complementarity and other exotic early twentieth century attempts to explain quantum mechanics.  

In general, there is no well defined ``location'' and using such language just takes us further from a language built around the persistent reality of a many body wavefunction as the fundamental descriptor of the system.  A better phraseology is to say that the coordinates of the wavefunction corresponds to various particle types and spin labels and this wavefunction has various preserved symmetries.  In some cases, the wavefunction seems to correspond to a set of sharply defined atomic locations.  This is a property we expect of classical condensed matter, especially solids.  

Let us consider the ground state of N atoms of an alkali metal.  It is natural to expect it to decompose into CM and internal coordinate parts with a two parameter rotational freedom in the internal part: $\Psi=\psi_{cm}(X_{cm})\psi_{int}(x^{rel}_{1}\ldots x^{rel}_{n-1})$ where there is a (passive) transformation $T_{(\theta,\phi)}$ of the $Y_{rel}=\{x^{rel}_{i}\}$ such that $\psi_{int}(Y_{rel})=\psi_{int}(T_{(\theta,\phi)}Y_{rel})$.  Colloidal condensation of metals gives special polyhedra based on the number of atoms they contain.  As this number grows they tend to spherical bodies yet the crystalline structure still destroys rotational symmetry.  We expect the one body density matrix of the true ground state to be roughly spherical.  Of course, the actual possibilities for solids are highly varied in shape, internal crystalline structure, nanoscale order and orientation.  The fact that our observation of solids show a particular shape, position, orientation and microcrystalline structure have both a well defined location and orientation suggests that the center of mass is localized and the rotational and many other internal degrees of freedom are in a highly excited state that allows particle locations to be well defined.  

Beyond this, the internal structure is likely in an internal superposition as well.  The apparent ``ground state'' of such a body is often associated with zero temperature but clearly requires many extra nodes corresponding to excitations and many body currents associated with being far from a true eigenstate.  When we investigate the primitive excitations of such a body we should to remember that there is this noisy hidden background.  The apparent excited eigenstates of these N atoms system will involve nontrivial combinations of the true eigenstates that may include bound and unbound states, ionization of some atoms and states with unclear surface boundaries in the one body density function and so on.  Such situations are exactly what we do not see in classical matter.  Justifying this gulf in observed behavior in terms of  wavefunctions is a task that should be considered of comparable importance to other fundamental problems such as the ``arrow of time.''  Here I will show it is not that hard, In fact, the resolution can be thought of as already implicit in some solid state and transport calculations.  

Nowadays, often some notions from decoherence are used to explain away why we don't see superpositions of macroscopic objects.  This is ultimately not much more satisfying since it often just seems like another act of modifying language with few (if any) testable consequences.   Instead of this we will proceed with a description of the world in terms of wavefunction consistent with our observations the classical world.  The large mass of a classically sized body means that the delocalization time for its position and orientation is enormous.  (Gases do not suffer such a constraint since the many body wavefunction can evolve freely in each coordinate most of the time.)  

This leads us to our first fundamental assumption about the quantum state of classical matter:

{\bf Proposition A}: Solids and liquids (with the possible exception of Helium) are described by long lived far-from-eigenstate wavefunctions where the atomic locations are well defined.  Specifically, the wavefunction is well approximated by an appropriately symmetrized products of delta functions in the atomic coordinate labels and symmetrized products of atomic orbitals in the electron coordinates located about the corresponding atomic orbital locations $\{R_{i}\}$.  The atomic coordinate part we will call $\Psi_{core}$ and the electronic part that has been projected out for a fixed value of the atomic locations at a local maximum is $\Psi_{e}$.  The appropriately symmetrized product of these we will call $\Psi_{class}$ which is a near approximation to the true wavefunction descriptor of the system.    
\begin{align}\label{class}
\Psi_{class}=\hat{\mathcal{S}}_{e}^{f}\hat{\mathcal{S}}_{a}^{f/b}\prod_{j=1}^{N} \psi_{0}(X_{j}-R_{j})\prod_{k=1}^{N}\phi(x_{k}-X_{j})=\Psi_{core}\Psi_{e}
\end{align}
where $R_{j}$ are the lattice locations of the atoms (including its nonchemically active electron coordinates explicity), $\psi_{0}(y)$ is a narrowly peaked function (meaning much narrower than the electron orbitals) indicating its position, $\phi(x)$ is the symmetry adapted electron orbital wavefunctions of the chemically bonding outer electrons.  $\hat{\mathcal{S}}_{e}^{f}$ is the (anti)symmetrizing operator for the electrons (with coordinates$\{x_{i}\}$).  
$\hat{\mathcal{S}}_{a}^{f/b}$ is the appropriate symmetrization operator for the atomic cores (coordinates $\{X_{j}\}$) depending on their Fermi or Bose nature.\footnote{Interestingly, this is pretty much the only sort of wavefunction you can write down that has the observed classical properties and degrees of freedom.  The innovative part is realizing that it persists for very long times and that we should leave the notion of eigenstates behind.  Given that measurement theory and quantum statistical mechanics are built on the notion of eigenstates, this may be a leap.  It requires one now reconsider both of these from another point of view as we do next.  }  
Spin labels have been suppressed but their effects are important in the symmetrization over ``particle labels''; specifically exchanges of the coordinate and spin labels of a particle together.

Since we have been discussing the confusing distinction between location and coordinate label this is a good moment to discuss the notion of discrete symmetry of a solids.  We know from STM images that we can ``see'' atoms at precise locations on a solid surface.  Above we see that the wavefunction of such a state encodes this in a very nonlocal fashion.   Often it is suggested that liquids have a continuous symmetry that is broken in the transition to the discrete symmetry of a solid.  Aside from the obvious problem that there are amorphous solids with no discrete symmetries even on very small scales, we see that while there can be a discrete set of symmetries among the lattice sites $(R_{1},R_{2}\ldots R_{N})$ when viewed in $\mathbb{R}^{3}$ such a symmetry is not evident in the wavefunction $\Psi_{class}$.  The CM is localized and so translations, continuous or discrete, do not map it into itself.  We can find a symmetry among the maxima when viewed in $\mathbb{R}^{3}$ but this is a model dependent statement.   
If we consider two lattices in $\mathbb{R}^{3}$ that differ in the number of particles as N and N+1 but otherwise overlap, the corresponding wavefunctions have different numbers of coordinate variables.  If we consider discrete translations of $\Psi_{class}^{N}$ along its CM by one lattice site in all coordinates then take a slice of $\Psi_{class}^{N+1}$ by fixing $x_{N+1}$ we see that we have a correspondence of the functions in that the maxima of $\Psi_{class}^{N}$
are found among those of $\Psi_{class}^{N+1}$.  In general, it is not possible to give a  discrete set of transformations for a wavefunction of a solid that correspond to those of the discrete 3D lattice associated with the core locations.  

Liquids have been a difficult phase to explain from a microscopic point of view.  Landau has stated \cite{Landau:smII} that the liquids have a continuous symmetry and therefore the ``phonons,'' as in the case of Helium, carry a mass flux and that this lets them contribute to the viscosity (as opposed to the quasimomentum of phonons in solids).  The inconsistency with this point of view is that such a flux also implies a high diffusion rate as in the case of gases.  The diffusion rate of liquids is far too small to justify this opinion.  Solids maintain shear stress by deformation of the atomic orbitals.  It seems likely that liquids do as well but allow for relaxation and this provides the microscopic mechanism for transfer of forces and dissipation.  This should be detectable spectroscopically.  Indeed for polymer melts polarization is observed under shear implying a general change in the microscopic order.  

This model of solids suggest that liquids (with the possible exception of Helium) have a similar localization property to solids but with an interconnection of potential wells that allows current to travel between nearby maxima.  If this is a good model one has to wonder if MD computations are missing an essential quantum feature in the description of liquids.  Electron clouds also become thermally excited making the Born-Oppenheimer approach potentially valid only if it is built from a quasistatic change of this excited electronic state.  It is unclear what a ``continuous translational symmetry'' means in such a model.  As we saw above, the discrete translational symmetry of a crystal only has meaning among the equilibrium core set $\{R_{i}\}$ when veiwed in $\mathbb{R}^{3}$.  The discrete translations don't have any obvious meaning in terms of the true many body wavefunction.  

We are now in a position to give an estimate to how much of an excitation energy must be possessed such a solid block due to its classical state.  We can take such a block and cool it to absolute zero but the localization energy still persists.  These will induce oscillations and currents in various correlated directions.  Certainly there is some metastability in the nuclear components.  This will create a high frequency small amplitude we will ignore.  Similar consideration goes the chemical structure of the material.  The true ground state (modulo those effects) will be rotationally symmetric not just in its shape but in the angular localization of the atoms.  We can estimate the first contribution from the surface area of the block.  If the energy per bond is $E_{b}$ we have that the energy contribution is $ U_{surf}\sim E_{b}L^{2}$  The angular excitation must localize the orientation to well within an atomic radii $d$.  This corresponds an energy of $U_{r}\sim \frac{\hbar^{2}}{M d^{2}}$.  This is made up of counter moving rotational modes.  Each one gives some radial excitation of the body.  Assuming the Young's modulus is $Y$ we have $U_{el}\sim \frac{\hbar^{2}}{MY d^{2} L^{3}}$.  These generate oscillations that are present regardless of how far the object is cooled and gives a measure of how far a classical object must be energetically from the true ground state of the system.  To ``cool'' the system further, the atoms must diffuse to a more symmetrical shape and the shape itself must delocalize into a state where the crystal locations are ambiguous.  Here it is even more evident that there is no fundamental translation corresponding to the discrete symmetry of crystal as an action on its many body wavefunction. 

\subsection{Phonons}
The description of the low order excitations of solids is typically done by the Debye model of examining the classical (longitudinal) normal modes and quantizing them.  The second quantized formulation is often introduced here in analogy with photons.  This distinction is that photons are actually elementary particles that introduce their own coordinate labels and so increase the dimensionality of the system.  Phonons exist as modifications of the solid 3N-D wavefunction (the ``pseudo-ground state'') and so we should be able to write down not a wavefunction for them, but describe them as a modification of the lowest energy wavefunction of the solid.  

Using our above wavefunction, we expect the atomic peaks to be bound in potential wells that are well approximated by harmonic potentials.  Shifting the peak locations gives linear restoring forces corresponding to the normal modes of the classical solid.  We have, of course, obscured the low amplitude many body currents hidden by our classical approximation of the wavefunction or, more accurately, our wavefunction description of classical matter.  
The modes generated by this many body potential well are the ``pseudo-eigenstates'' of the wavefunction, specifically long lived nearly orthonormal excitations that we suppose dominate the energy of thermal motions.  Naturally these correspond to phonons.  Consider the one \textit{particular} peak of $\Psi_{class}$ at $\tilde{R}^{(1)}=(X_{1}=R_{1},X_{2}=R_{2}\dots X_{N}=R_{N})$ and displace by $\delta x$ in all directions to obtain the effective many body harmonic potential.  Let the 3N-5 directions 
of the eigenstate of the $i$th mode be indicated by $k^{(i)}_{l}=(y_{1},y_{2}\ldots y_{3N-5})^{(i)}$ where $k^{(i)}$ is normalized to one, $i$ indicates the mode type and $l$ the particle coordinate.  This gives frequencies $\omega_{i}$ and effective masses $m_{i}$.  The corresponding eigenfunctions for the $n$th excitation level is 
$\psi_{n_{i}}(x)=f_{n}(m_{i},\omega_{i})H_{n}(\sqrt{\frac{m\omega_{i}}{\hbar}}x)$.  These can be though of as centered about our particular $R_{i}$ peak and oriented along $k^{(i)}$.  Putting these all together to obtain the wavefunction of such a solid with excitation level $\{n_{i}\}\in\mathbb{N}^{3N-5}$ in each of these 3N-5 modes: 
\begin{align}\label{phonon}
\Psi_{class}(\tilde{X},\tilde{x})=\hat{\mathcal{S}}_{e}^{f}\hat{\mathcal{S}}_{a}^{f/b}\prod_{i}^{3N-5}\psi_{n_{i}}\left(\sum_{j}(X_{j}-R_{j})k^{(i)}_{j}\right)
\psi_{cm}(X_{cm})\psi_{r}(\theta,\phi)\prod_{k}^{N}\phi(x_{k}-X_{j})
\end{align}
where $\psi_{cm}$ is a narrow function.  To be consistent with the originally proposed $\Psi_{class}$ we should choose its width to be that of $\psi_{0}$.  The rotational function $\psi_{r}$ specifies a particular narrow angle fixing the orientation of the body.  These two functions will be narrow but long lasting nearly monochromatic packets if the body is translating or rotating.  The peaks are centered so as to fix the last five $R_{j}$'s.  
This follows from our expectation that these are gaussian like peaks in each coordinate so hyperspherically symmetric in the atomic core sector of the many body space.  
The oscillations of the core locations give restoring forces from the shifted atomic orbitals.  This arises because the energy of the overlapping $\phi$ change as a function of relative position of the cores.  Our assumption here is that this is not enough to appreciably change the from of the atomic orbitals used in this approximation.

In nuclear physics there is an analogous problem for highly deformed nuclei.  Both mean field and Slater determinant approaches violate the rotational symmetry of the ground state.  This can be put back in \cite{Yoccoz} by introducing rotational symmetry of such a state.  For strongly deformed nuclei, this makes a difference in the final energies.  For our localization functions $\psi_{cm}$ and $\psi_{r}$ this makes an extremely small difference in the energy as estimated in Sec.\ \ref{solids}.  However, the state is at best metastable (a pseudo-ground state) arrangement so the energy of this state is actually appreciably far from the true ground state.  Furthermore, we shall see that it makes a huge effect on the dynamics of the system as it interacts with delocalized particles and gives a local and satisfying explanation of quantum measurement.

In the usual phonon language we would say that the $i$th mode has occupancy $n_{i}$.  Here we see that this corresponds to a particular many body direction having where the wavefunction has $n_{i}$ oscillation nodes.  If we were to excite to higher levels some anharmonicity would show up in the form of these potentials and their general separability would fail.  This might be mistaken to mean that a nonlinear interaction is causing phonons to scatter.  However, it only means that the many body pseudo-eigenstates now have excitations that don't increase by constant energy steps and the net many body states don't separate.  Second quantization is built around such harmonic states so it is hard to glean this geometric intuition from it.  Typically there are phonon interactions or other more ad hoc approaches.  Some of these may be valid but it would be an advantage to have a geometric description, and its associated improvement in intuition, to suggest which ones.  This will be discussed further in Sec.\ \ref{Fock}.

We can compare the widths of the relative ground state $\psi_{n}$ functions.  The widths $w_{i}\sim ({m_{i}\omega_{i}})^{-1/2}=(\kappa_{i} m_{i})^{-1/4}$ where $\kappa_{i}$ is the effective spring constant of a given mode.  Since the spring constant doubles when we half the wavelength of a mode and this reduces by half the amount of mass involved we see the widths are invariant.  This gives a consistency check that our ground state function in Eqn.\ \ref{phonon} has the same localization as in Eqn.\ \ref{class}.  In computing the energy of these excited states we should note that the localization given by $\psi_{0}$ was optimized to give a minimum.  This implies that we do not need to include a zero point energy contribution per mode.  It was already included in the energy.  If we do include it, the potential energy must be combined with it to give an identical net energy of the system.  This has obvious implications for any theory of the Casimir effect based on phonons.  

It is interesting to consider what happens in the case of Fermionic atom cores.  These will still localize to give peaks at the many body points given by all permutations of $\tilde{R}=(R_{1},R_{2}\ldots R_{N})$.  The space is now divided into a set of nodal cells.  In dimensions higher than one, this is typically two or four \cite{Ceperley:1991} for the ground state.  The peaks now divide into equal sets of positive and negative amplitude but they still have the same decomposition into a set of normal modes.  These give a set of 3N-3 directions that can each be excited to increasing numbers of oscillations.  This is the reason that phonons obey Bose statistics when enumerated in second quantization even for fermionic atoms.  The actual wavefunction does not represent phonons as anything but nodal bifurcations of the localized peaks corresponding to a classical solid's atomic locations.

Despite formal similarities, phonons are very distinct from actual material particles.  As quantized sound modes they transport no mass hence have true momentum density of zero.  They are labeled by $k$-vectors corresponding to the oscillation frequency given by the normal mode decomposition of the many body potential for the long lived localize states at the N! maxima in the $\mathbb{R}^{3N}$ configuration space.  There is a long confusion about the distinction between true momentum, pseudo-momentum and crystal momentum \cite{Mc81}.  Most solid state books give a brief mention of the distinction between between the latter two but often mislabeling the pseudo-momentum as ``true momentum''.  Enduring confusion persists because of the coupling rules in field theory methods that ``conserve (pseudo)momentum'' without mentioning that it is pseudo-momentum.  A similar situation has occurred in acoustic, plasma physics and the theory of ocean waves.  Often in places where some field theory method or a heavy vector calculus treatment of a hydrodynamic system is used.  The confusion often begins with the choice to fold in the nonlinear advection term $v\cdot\nabla v$ into the stress term $\Pi_{ij}$.  
\begin{align}
\rho\partial_{t}v+\rho v\cdot\nabla v&=-\nabla\cdot\Pi\\
\partial_{t}(\rho v)&=-\nabla\cdot (\Pi+\rho v\varotimes v)=-\nabla\cdot\Pi'
\end{align}
This is mathematically and dimensionally consistent but is properly a ``pseudo-stress'' with no microscopic representation as true stress hence of dubious value in deriving true forces on the boundary of or within the medium.  
It is clearly understood that the group velocity is the relevant quantity to discuss the transport of mass hence specify momentum (since ``mass flux''=``momentum density'' for massive particles except for tiny relativistic corrections) in the semiclassical theory of electrons \cite{Ashcroft} however in the case of phonons there is no mass transport at all.  The situation grows even more confusing for the case of superfluid Helium where the linear part of the dispersion relation is labelled ``phonon'' and mass is actually transported over significant distances with these modes \cite{Khalatnikov:65}.  

Inspecting the origin of momentum conservation rules for Feynman diagrams in QED we see that they arise from the Fourier transform of the spatial locality condition on the interaction.  Conservation laws alone can dictate conservation of momentum order by order in the Dyson series but not diagram by diagram.  Transporting these ideas to condensed matter theory suggests it is a good idea to conserve such pseudomomenta in diagrams as long as the true momenta is conserved as well.  This is a bit of a subtle question and we won't pursue it further here.  

\subsection{Fock Space and Quasiparticles}\label{Fock}

One can give a many particle description of a quantum system in terms of Fock space.  This is generally accomplished by the use of creation and annihilation operators and a formal algebra.  Since we are interested in geometric intuition and consistency, we present and explicit model for such actions here.  Firstly, for the case of true particles then in the case of quasiparticles corresponding to collective modes.  

One can write a many body space representation for spinless bosons as $\Psi=a+b\bra{\Psi^{(1)}(x_{1})}+c\bra{\Psi^{(2)}(x_{1},x_{2})}\ldots $ with appropriate symmetries for each wavefunction.\footnote{We are included a ``vacuum state'' constant value here.  When we consider phonons, we will revisit this.}  In the case of second quantized treatments, the dominant way to treat such systems when particle numbers can change, we are interested in using creation and annihilation operators with respect to a particular one-body basis $\{\psi_{i}(x)\}$.  In this case it is good to write the basis of each N-body function explicitly as a symmetrized product on the basis 
\begin{align}
\{1,\{\psi_{i}(x_{1})\}, \mathcal{\hat{S}}\{\psi_{i}(x_{1})\psi_{j}(x_{2}) \},\mathcal{\hat{S}}\{\psi_{i}(x_{1})\psi_{j}(x_{2})\psi_{k}(x_{3})  \}\ldots    \}=\{1,\psi^{(1)}_{i},\psi^{(2)}_{ij},\psi^{(3)}_{ijk}\ldots\}
\end{align}
where $i\le j \le k \ldots$(and all other such $\psi_{ijk\ldots}$ are taken to be zero).\footnote{Typically, the coefficient sets are limited in such expansion as a consequence of symmetry.  Here we have restricted the basis set instead so that any other coefficients not satisfying this index constraint have no effect.}  A general object is then representable as 
\begin{equation}
\Psi=\sum_{N=0}^{\infty} \sum_{i_{k}=1}^{N} b^{(N)}_{i_{1},i_{2}\ldots i_{N}}\psi^{(N)}_{i_{1},i_{2}\ldots i_{N}}
\end{equation}
A creation operator $\hat{a}_{s}^{\dagger}$ can then act on $\Psi$ in this representation by acting on the coefficients.  From this example the most natural definition would be to define the map 
\begin{align}
\hat{a}_{s}^{\dagger}: b^{(N)}_{i_{1},i_{2}\ldots i_{N}}\rightarrow \begin{cases}b^{(N+1)}_{P(i_{1},i_{2}\ldots i_{N},s)}&\mbox{if} ~N>0\\
0  &\mbox{if} ~ N<0\end{cases}
\end{align}  
where $P$ induces the above partial ordering among indices and the map gives the shift of the values of the coefficients in the expansion.  This operator is well-defined since it is one-to-one.  
The corresponding annihilation operator would be
\begin{align}
\hat{a}_{s}: b^{(N)}_{i_{1},i_{2}\ldots i_{N}}\rightarrow \begin{cases}b^{(N-1)}_{i_{1}\ldots\hat{s}\dots i_{N}}  &\mbox{if}~ s\in\{i_{k}\}\\
0&\mbox{if}~s\notin\{i_{k}\}\end{cases}
\end{align}  
and the $\hat{s}$ indicates that (one of) the $i_{j}=s$ label(s) has been omitted.  In practice these operators are defined with extra coefficients that are functions of the ``occupancy'' of the state, specifically the number of $s$-indices present in the coefficient.  Let $n_{s}$ be the number of such indices in the $\psi^{(N)}_{ijk\ldots}$ coefficent $b^{(N)}_{ijk\ldots}$.  By letting
\begin{align}
\hat{a}_{s}^{\dagger}&\rightarrow \sqrt{n_{s}+1}~\hat{a}_{s}^{\dagger}\\
\hat{a}_{s}&\rightarrow\sqrt{n_{s}}~\hat{a}_{s}
\end{align}
we can get the commutation relations $[\hat{a}_{s},\hat{a}_{t}^{\dagger}]=\delta_{s,t}$ and  the identity for the number operator for the occupancy $n_{s}=\hat{a}_{s}^{\dagger}\hat{a}_{s}$.   In the case of fermions we can construct a similar basis with introduction of spin labels and restriction on occupancy due to antisymmetry.  The manifest representation of creation and annihilation operators then gives a set of anticommutation relations, as familiar in all quantum textbooks.  

Typically calculations of this sort are done strictly formally.  In the case of fixed N-body quantum mechanics in an external potential, this can be shown to be equivalent to the usual many body Schr\"{o}dinger equation evolution \cite{Schweber}.  The downside of this approach, aside from its formality, is how to describe this reality in a manifestly basis independent fashion?  One can transform to other one-body basis sets in an obvious manner but this still not completely general.  In principle one could use two-body bases but a truly geometric discussion would be more along the lines of the Schr\"{o}dinger equation itself where there is no decomposition in terms of a basis and explicit manipulation of coefficients arises in the description.  In the case of real particles, it is not immediately clear how to proceed but in the case of collective modes (phonons) we will show the situation is a little better.  

In the case of phonon ``creation'' and ``destruction,'' note that this does not change the number of coordinate variables at all.  Specifically, if the system is described by N coordinate labels at one time, $\Psi^{(N)}(x_{1},x_{2}\ldots x_{N})$, it is so at all times.  To describe the action of these operators on one of the basis functions
\begin{align}
\Psi_{core}(\tilde{X};\{n_{i}\})=\hat{\mathcal{S}}^{f/b}\prod_{i}^{3N-5}\psi_{n_{i}}\left(\sum_{j}(X_{j}-R_{j})k^{(i)}_{j}\right)
\psi_{cm}(X_{cm})\psi_{r}(\theta,\phi)=\hat{\mathcal{S}}^{f/b}\Phi(\tilde{X};\{n_{i}\})
\end{align}
we need a way to shift the occupancy of a particular mode.  We can think of $\Phi(\tilde{X};\{n_{i}\})$ as the local function of a solid with phonon occupancy $n_{i}$ about the many body location specified by the lattice positions in their ``standard order'': $\{R_{1},R_{2}\ldots R_{N}\}$.  Seeking a kernel for $\Phi$ alone to raise the occupancy of the the $k^{(s)}$ phonon from $n_{s}$ to $n_{s}+1$\begin{align}
\tilde{K}_{n_{s},n_{s}+1}^{(\Phi)}(\tilde{X},\tilde{X'})=K_{n_{s},n_{s}+1}((X_{j}-R_{j})k^{(s)}_{j},(X'_{j}-R_{j})k^{(s)}_{j})T(\{(X_{j}-R_{j})k^{(i)}_{j}\},\{(X'_{j}-R_{j})k^{(i)}_{j}\})
\end{align}
 where $T(\tilde{X}_{\perp},\tilde{X'}_{\perp})$ is a transverse 2(3N-3)-D function (so uses all $i\ne s$) that maps all (N-1)-fold products of phonon oscillations (and the five CM and angular localization functions) into themselves.\footnote{$K_{n,n+1}$ here really only needs one subscript since we are not defining it for all subscripts in $K_{ij}$.  The pair is just a reminder of the action.}
  $\tilde{K}^{(\Phi)}_{n,n+1}$ is chosen so that
\begin{align}
\int d\tilde{X'} \tilde{K}_{n_{s},n_{s}+1}^{(\Phi)}(\tilde{X},\tilde{X'})\Phi(\tilde{X};\{n_{i}\})=\Phi(\tilde{X};\{n_{1}\ldots n_{s}+1\ldots n_{N}\})
\end{align}

Since these peaks are very weakly overlapping, even for high phonon occupancy, we can define a kernel 
\begin{align}
\tilde{K}_{n_{s},n_{s}+1}(\tilde{X},\tilde{X'})=\hat{\mathcal{S}}^{b}\tilde{K}_{n_{s},n_{s}+1}^{(\Phi)}(\tilde{X},\tilde{X'})
\end{align}
that performs the same action on $\Psi_{core}(\tilde{X};\{n_{i}\})$.  The corresponding form of the creation and annihilation operators are:
\begin{align}
\hat{a}_{s}^{\dagger}&=\sqrt{n_{s}+1}~\int d\tilde{X'} \tilde{K}_{n_{s},n_{s}+1}(\tilde{X},\tilde{X'})\\
\hat{a}_{s}&=\sqrt{n_{s}}~\int d\tilde{X'} \tilde{K}_{n_{s},n_{s}-1}(\tilde{X},\tilde{X'})
\end{align}
where it is understood that a function $\Psi(\tilde{X'})$ stands to the right inside the integral.  
Note here that the operators are only defined for integers $n_{s}>0$.  We can define the action of $\hat{a}_{s}$ on occupancies $n_{s}=0$ (i.e.\ the ground state) to be zero.  This completes the operators definition and ensures the usual commutation relations.  Unfortunately, the operators require a function of well define $n_{s}$ on the RHS so it is not suitable to give a manifestly basis-independent description for any Hamiltonian built from such operators.  It is possible that the kernel could be modified to be invariant for all pairs of consecutive integer indices but the coefficient out front spoils any possibility of such a construction.  If we abandon this coefficient, as in our initial guess above, we lose nontrivial commutation relations but might obtain a basis invariant form of the equations of motion that map onto the usual second quantized approach.  

The question of locality is still somewhat vexing.  For the many body Schr\"{o}dinger equation, locality is manifest and the natural basis for second quantization for such a description is the $\delta$-function set.  In terms of the collective modes, the operators are integrals so quite nonlocal.  To understand if locality is really preserved we should consider the way that these operators couple modes together and what is the microscopic origin of these rules.  To do so, we should consider the kinds of models used that cause phonon interactions.  Heat transport and electron-phonon interactions are often framed in the momentum representation.  As in the case of QED, the spatial locality of the interaction fixes (pseudo)-momentum conserving relations among the diagrams.

Acoustic phonons are often the canonical example of quasiparticles.  Some quasiparticles, like electrons, can carry mass so real momentum as well.  The discussion of ``dressed electrons'' from this point of view would get rather lengthy but we can now suggest what sort of many body systems are amenable to a quasiparticle treatment.  In the one particle case, we can turn up a potential and observe the changes in the energy levels.  Typically there are avoidances but some symmetries allow (generally transverse) crossings to occur.  This allows one to map the original $\ket{k}$ labels of the free eigenstates to the ones of those with interactions.  

In the many particle case, interactions can be viewed as complicated sort of potential.  The free body states $\bra{k_{1},k_{2}\ldots k_{N}}$ map, as above, to new interacting states.  However, we still would like to use the one-particle picture as much as possible.  Generally, the interactions destroy the symmetrized product form of the eigenstates so we need to know how to give meaning to a state $\ket{k}$ with such strong interactions (and when it has any meaning at all).  Based on the generally desired properties of quasiparticles, additivity, we can suggest that the low energy many body eigenstates be of the form:
\begin{equation}
\Psi_{n_{1},n_{2},\ldots n_{N}}(\tilde{X})=\hat{\mathcal{S}}^{f/b} \int d\tilde{u} \prod_{i} f_{n_{i}}(u_{i})F(\tilde{X},u_{1},u_{2}\ldots u_{N})
\end{equation}
The coordinates $u_{i}(\tilde{X})$ give a tailored set of coordinates to orient to the optimal independent low energy oscillations of the ground state.  For phonons $u_{i}=(X_{j}-R_{j})k^{(i)}_{j}$.  The kernel $F$ acts similarly to a Jastrow correction by introducing potentially strong N-body changes to the ground state while the $f_{n}$ functions give the oscillations along these optimal directions.  To the extent that such a wavefunction holds, we can think of the system as made up of quasiparticles.  Notice that we have not included a CM or angular localization function.  Classical bodies will require this but we are also interested in systems like gases, ultracold gases, and electron systems where such a localization may not be present.  In the absence of such localization, this is now a set of true low energy eigenstates.

In the case of a solid, when it heats to the point that is starts to suffer a net expansion (so the core locations shift) or anharmonicity arises in each of the optimal many body directions, this picture starts to fail.  Thermal conductivity models incorporate such changes while retaining the phonon picture by introducing classical kinetic notions such as ``mean free path'' and ``collision time'' and by introducing coupling between phonons.  The first set of notions is troubling because the phonons are intrinsically nonlocal objects that don't form packets in any obvious way.\footnote{The closest approximation to a phonon packet is probably as follows.  Consider a connected subset of the core locations $\tilde{R}'\subset \tilde{R}$ and define a corresponding basis of phonon oscillation functions corresponding to these.  The resulting set of states can be expanded as functions on $\tilde{X}$ by including localization functions in the unused directions.  These must then be expressible on the original phonon basis $\psi_{n_{i}}(\tilde{X})$.  The resulting collection of modes are initially localized in space and restricted to the frequency set corresponding to the smaller volume.}  The second is troubling because, anharmonicity just indicates that our harmonic approximation of the modes is no longer valid.  These models have met with some success though there is certainly continued debate about their success and validity.  Hopefully, this perspective can shed light on the validity of green's function methods and how to best modify them in such situations.

\subsection{Newtonian Limit}
The task of deriving classical mechanics from quantum mechanics has seemed intractable for many reasons.  Ehrenfest (Ehrenfest) offered a naive result for a localized packet and showed that given and arbitrary external electromagnetic field with gradients small compared to the packet localization one recovers the classical trajectory.  The larger the mass of the object, the longer the packet stays localized and the result is convincing.  For large mass objects, like all objects we experience through our senses, this seems plausible, at least for the CM coordinates.  One problem is that there is no reason that nature has to provide us with only localized large mass bodies much less ones with well defined shape and internal structure.  Additional complications like the ``Schr\"{o}dinger's cat'' and ``Wigner's friend'' paradoxes \cite{Jammer} leave us with macroscopic superpositions that seem quite unlike the observed natural world.  

By assuming that all macroscopic matter starts in the state described by Assumption~{A}, we have a chance of getting such a limit and we will see shortly that it can be greatly expanded.  
We have no immediate reason for starting with such a state based on assumptions about the state of the early universe, only that it gives the proper kinematic features over long times.  
As a first step in such a ``Quantum-Classical unification''  we should at least show that a rigid body, as we described by $\Psi_{class}$, moves freely according to Newton's laws.  Ehrenfest's Theorem demonstrates the center of mass motion moves properly.  The rotational motion part is more subtle.  Rotation of electrons about atoms is irrotational and introduces radial oscillation and the characteristic non-rigid body rotation.  These are wavefunctions as well so rotation requires singular vortiticy.  However, this can be outside the primary support of the body.  For a solid, internal deformations are energetically expensive.  The localization in the primary angular direction means there are large regions of low amplitude for vorticity to occupy unlike for the true rotationally symmetric ground state.\footnote{Ultracold gases in spherical traps have an energy cost for vorticity penetration and the low energy excited states are, accordingly, propagating irrotational surface waves with singular vorticity near the surface.}  

This implies that rigid body rotation follows by conservation of angular momentum and that all deformations of this function for a solid have high energy cost.  We can compare the kinetic energy of a rotating body to its binding energy.  $\frac{K}{B}\sim \frac{mv^{2}}{u}$ where $m$ is the mass of a particle and $u$ is the energy per bond.  In any angular momentum conserving scenario we expect deviations to have to overwhelm the binding energy.  This is approximately when the edge of the body approaches the speed of sound.  A similar argument explains why the internal structure is so slightly affected by rotation.  Centrifugal expansion can occur but the local relative positions of the cores remains largely unchanged.  In the true ground state and nearby eigenstates, such core locations are not well defined.  

\section{Measurement: Quantum-Classical Interactions}
\subsection{Overview}

The quantum theory of measurement, generally via the Born interpretation, consists of a quantum system to be measured, a macroscopic object that is classical and some experimentally determined property of it that measures a quantum state of the system.  The most basic of quantum measurements is that of position.  Measurements are presumed to correspond to a linear self-adjoint operator.  A canonical example is that of a delocalized atom, electron or photon incident on a screen.  $\rho(x)=\psi^{*}(x)\psi(x)$ gives the probability density that a the particle ``collapses'' at x.  

Some shortcomings of this point of view are as follows.  If the beam has cylindrical symmetry perpendicular to the screen this gives a unique result, however, one can certainly construct wavefunctions of a single particle that are much more varied.  For example, consider a wavefunction broken up into two spatially separated packets traveling at different velocities and directed at different areas of the screen.  The time of the the events must be part of what we ``measure'' and these we don't expect to occur at the same time.  Indeed the packet length can be much longer than any measured ``collapse time'' and the measurement device can sometimes accelerate and move faster than the wavefunction itself.  

We can construct correlated incident systems of N particles.  The above criterion tells us nothing about the expected probability of results in correlated measurements or if measurement devices can be constructed that look specifically at linear operators of two body systems.  Certainly, there has been much work on hidden variables \cite{EPR} \cite{Bell} that do look at correlated measurements but these require a way to apply our measurement devices into the larger dimensional space described by these functions.  As such, this is properly an extension of the original Born interpretation.  Certainly we should be able to take different measurements like the position of one particle and momentum of another.  In principle the granularity scale of measurement should be something that can be modified as well.  (An example could be a lattice of cold atom traps with large delocalization.)  Having a more intrinsic theory of measurement could suggest a broader range of questions and macroscopic interactions with delocalized bodies.  

The microscopic details of a body should say something about what is ``measures.''  The efficiency of the body to make a measurement versus reflect, allow transmission or otherwise interact with the body should also be something we can derive.  We already apply quantum ideas to describe microscopic properties of condensed matter.  It seems overly pessimistic to accept the positivist limitations on our ability to further apply quantum ideas to get a complete description of bulk matter.  The velocity of delocalized particles can be quite low and their packets quite extended.  We have the freedom to change location of the measurement device as it interacts with the particle.  This can be done in a noninertial fashion.  It can accelerate, rotate, even deform and change its chemical and electrical properties mid-measurement.  Somehow, the results of such experiments should be expressible in our theory.  Uncollected portions of the wavefunction can be altered by external fields during a later pass at it.  For continuous sources, it is often unclear how to think of the beam of particles emerging from it.  For example, are these correlated?  How large is the wavepacket for each one?  If one uses a ``chopper'' to partition the beam with a unit mass, does this correspond to one particle or partial combinations of several particles where the rest of their amplitude has been averted?  

All these concepts suggest that there is a kind of ``incompleteness'' to quantum mechanics as it currently stands.  This is not the incompleteness suggested by Einstein, Podolsky and Rosen \cite{EPR} where hidden variables rescue determinism and locality.  Rather, it is a incompleteness in doing the basic job of giving a full description of what we can measure even given a probabilistic interpretation.  Relativistic concerns are well known for the measurement problem.  The Newton-Wigner operator \cite{Jammer} is a (failed) attempt to extend the position measurement operator to the relativistic domain.  More generally, the instantaneous and delocalized nature of measurement seems hopelessly at odds with relativity.  

One can even argue that the basic Born interpretation is often not consistently applied in the cases where it is used.  The measurement operator $\hat{X}$ is a 3D operator that acts instantaneously everywhere.  (Presumably the rigid measurement device specifies a particular reference frame.)  In practice we make 2D measurements on screens or measure paths in cloud chambers or, their modern successors, semiconductor and wire arrays.  These give paths that could be viewed as a sequence of position measurements with lost time information or as momentum measurements.  The momentum eigenstates are properly delocalized in space so do not give a quasi-1D path.  Maybe the best interpretation would be to consider these as continuous measurement sequences that measure position in the transverse direction and momentum in the longitudinal direction.  Of course, the very notion of ``transverse'' is dependent on the choice of longitudinal one so these are not really independent measurements.  
Regardless of such ambiguity, physicists have found ways to apply these rules to pull out useful and accurate data and relate it to expressions in the theory.  

\subsection{Surface Position Measurement}

It was our supposition that quantum measurement is the result of a small delocalized or otherwise nonclassical body interacting with condensed matter, typically solids.  For a first special case let us consider our special case of a solid $\Psi_{class}$ with an extended flat suface normal to the direction of propagation of an incident particle.  We choose it at rest and investigate the action of an incident delocalized atom (distinct in type than that of the solid) that will bind chemically to its surface.  The oscillation energy density of the packet $\psi$ is assumed never great enough to penetrate the surface and the support of its width is less than that of the surface layer of atoms at the time of contact.  Furthermore, that the center of mass, angular and shape changing spreading rates of our solid body are much smaller than the temporal variations in energy, charge, momentum, etc.\ density of the incident particle wave.  (This is almost always satisfied.)   Chemical changes during binding are also assumed to occur rapidly compared to any oscillation time of such quantities and $\psi$ has slow spatial variations in these quantities on the granularity scale (atom size) of the body.  

At $t=0$ we can describe this as a product function $\Psi_{class}(\{x_{i}\},\{X_{j}\})\psi_{inc}(x)$.\footnote{The $x$ coordinate describes the CM location of the incident $\psi$.  The $X_{j}$ and $x_{i}$ coordinates are the core CM and electron coordinates of the body.  The electron coordinates of the incident $\psi$ have been suppressed.}     When the wave interacts with the surface we generally assume it adsorbs to single site.  The conservation laws of center of mass motion seem to be violated by this event.  If it were not, there would need to be a rapid impulse to shift the measurement device to correct for it.  Other conserved quantities have similar problems.  Everett \cite{Everett} proposed a kinds of many worlds interpretation that has the potential to solve this and was a favorite among many of the old guard looking for an ``interpretation'' of the wavefunction.  Instead of a radical transformation of the observed system, the universe, instead, obliges to bifurcate itself into the plethora of possible outcomes.  
We will see that something similar arises here in that the the space is partitioned into a distinct number of spaces determined by the granularity scale of the detector at every measurement event.  However, the key is just an implication of our particular subset of wavefunctions that describe macroscopic objects.   

\begin{figure}
\bc\includegraphics[width=10cm]{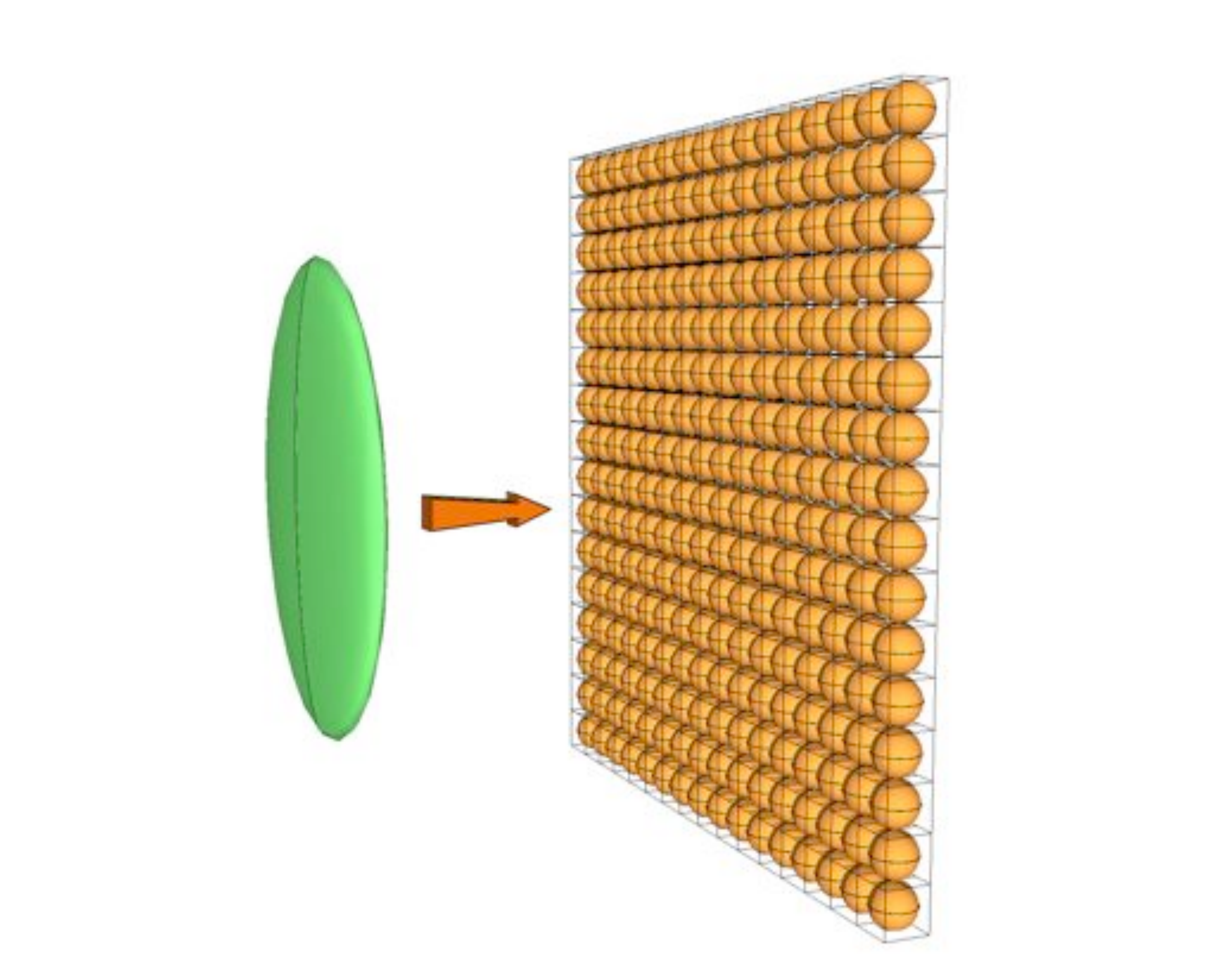}\ec
\caption{\label{fig:flat_incident} A particle with a broad narrow CM packet is incident on a surface.  The orange spheres evoke a sense of the electron cloud size although we are concerned here primarily with the localization of the cores (which are much more tightly localized).  The green oval indicates the dominant support of the incident packet and the orange arrow, the net direction of motion.} 
\end{figure}

\begin{figure}
\bc\includegraphics[width=10cm]{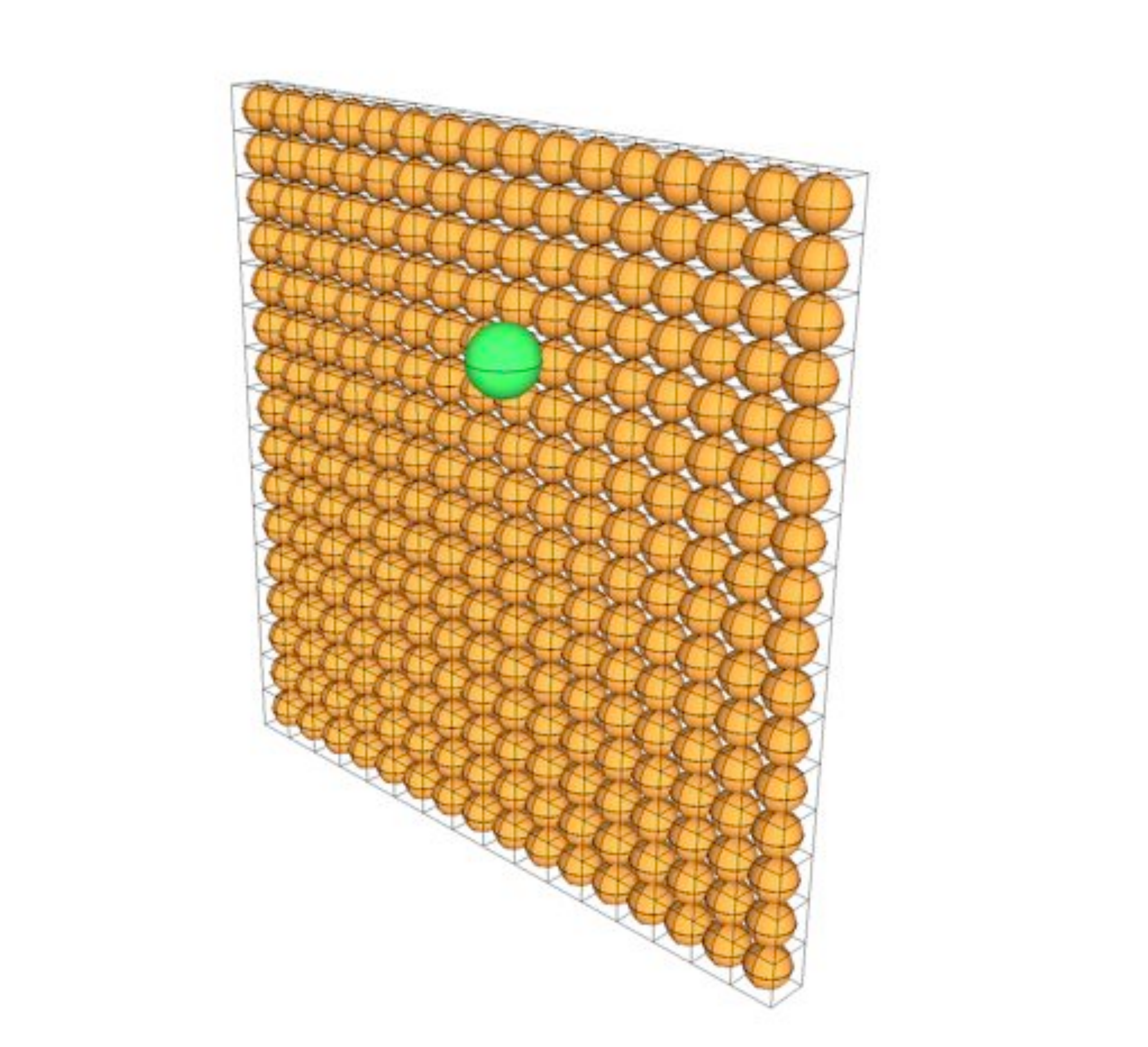}\ec
\caption{\label{fig:collapsed} The resulting state of the system in a slice where one location is selected.  }
\end{figure}

\begin{figure}
\bc\includegraphics[width=10cm]{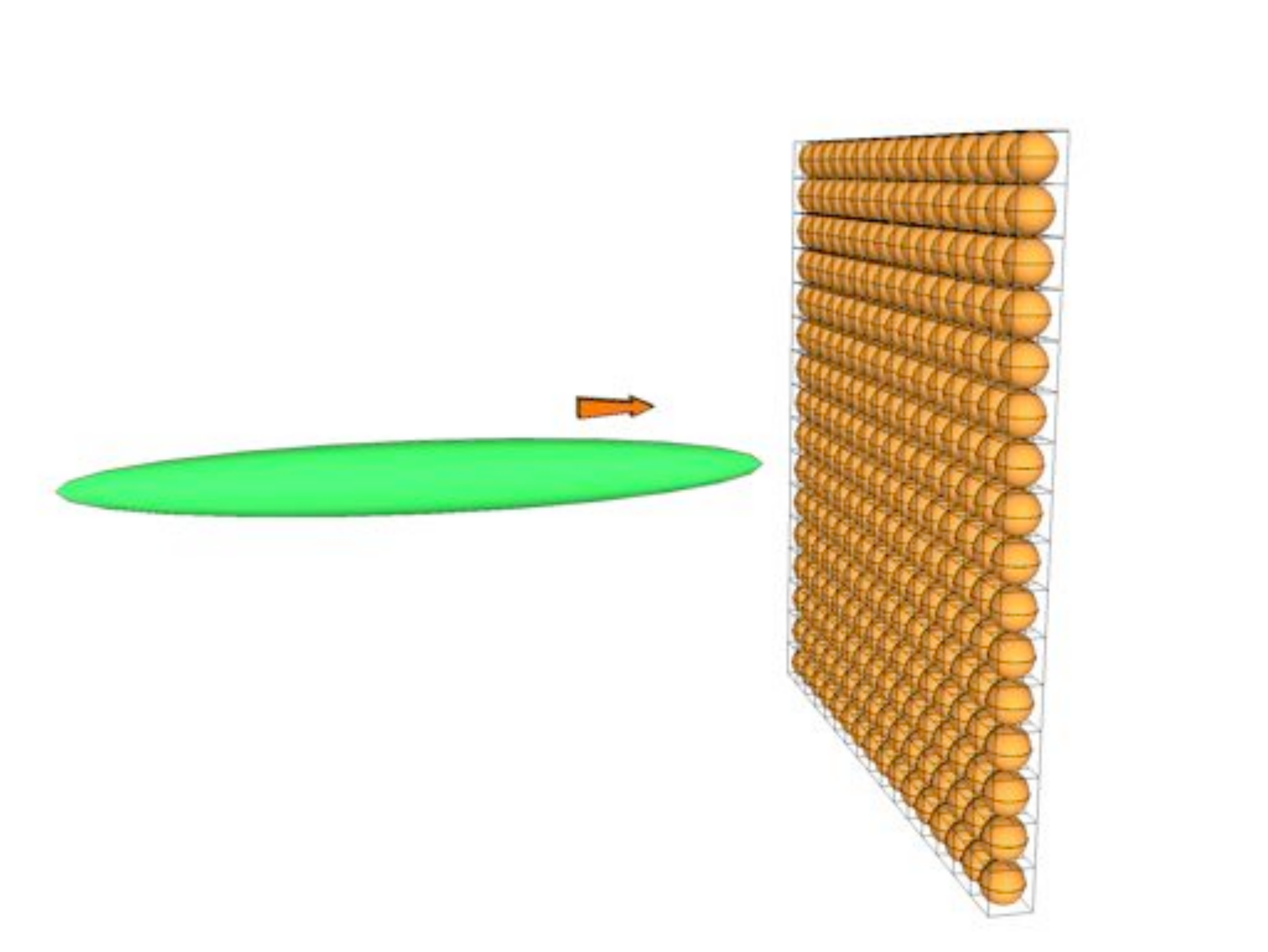}\ec
\caption{\label{fig:narrow_incident} A particle with a narrow CM packet that remains incident at the same location on the surface but that has amplitude arriving over an extended period of time. }
\end{figure}

For the moment, let the incident packet be a very narrow sheet that meets all points of the surface at the same time $t_{1}$ as in Fig.\ref{fig:flat_incident}.  The wavefunction forms bound states at each point and gives a new function (up to thermal oscillations and radiation loss contributions) 
\begin{align}
\Psi'(x,\{x_{i}\},\{X_{j}\},t_{1})=\hat{\mathcal{S}}\sum_{s}\psi(R^{(s)},t_{1})[b(x-R^{(s)})\Psi_{class}^{*}(\{x_{i}\},\{X_{j}\},k)]
\end{align}
where the bracketed fraction is the wavefunction with localization at a site and the left factor is the weighting factor that now parameterizes the state.  The local linear and angular momentum and energy on the granularity scale is transferred to the each corresponding site on the now \textit{excited} medium $\Psi^{*}$ that includes the acoustic recoil and localized heating of the adsorption.  
$R^{(s)}$ indicates the set of $N_{s}$ surface binding sites of the solid.  
The energy cost of localizing the incident wave on the granularity scale to the narrow packet $b(x)$ is given by the binding energy.  We see that $\psi(x,t_{1})$ acts as a parameterization for the set of $N_{s}$ solids with bound surface atom in the N+1 dimensional space (spin and other labels are suppressed).  
The result valid for all times is  
\begin{equation}
    \Psi(x,\{x_{i}\},\{X_{j}\},t)=
    \begin{cases}
      \hat{\mathcal{S}}\psi(x,t)\cdot \Psi_{class}(x,\{x_{i}\},\{X_{j}\},t), & \ t<t_{1} \\
      \hat{\mathcal{S}}\sum_{s}\psi(R^{(s)},t_{1})[b(x-R^{(s)})\Psi_{class}^{*}(\{x_{i}\},\{X_{j}\},t)], & t>t_{1}
    \end{cases}
  \end{equation}
The original $\psi(x)$ values play the role of introducing phase shifts and a weighting amplitude among similar copies of the original macroscopic body.  

We can now appeal to the strongly localized classical aspect of the original states.  Since the delocalization times of the solid are very long we can view the space have having been effectively ``sliced'' into $N_{s}$ copies that evolve as independent classical bodies one example of which is in Fig.\ref{fig:collapsed}.  For an observer or recording device riding along with the system, this generates the usual $\psi^{*}(x,t_{1})\psi(x,t_{1})$ probability distribution over many such events.  This situation will continue until the classical nature of our solid starts to fade, for example, when we have imparted so many delocalized bodies relative to the total mass of the measurement device that it has inherited a net center of mass delocalization from them.  Essential to this argument is that the CM localization of the solid is enough so that the slices from neighboring adsorption events do not overlap.  We define a ``slice'' to be a wavefunction describing the macroscopic body with an adsorbed particle at a specific site that occurred at a particular time $t'$.

Next we consider a very narrow but elongated incident $\psi$ packet traveling in a straight line (to the right along the x axis) with very little transverse spreading as in Fig.\ref{fig:narrow_incident}.  This strikes the same point $R^{(s)}_{k}$ but does so over an extended period of time from $[t_{0},t_{1}]$.  For simplicity, let us assume the surface is located at $x=0$ The set is now parameterized by $\psi(R^{(s)}_{k},t)$ where the value of $s$ determining the site is only one fixed value and $t$ is the time to get the value of $\psi$ at that location.  (We will integrate over the time of overlap.)  We already expect the final state to be a sum of terms like $\hat{\mathcal{S}}[\psi(R^{(s)},t')b(x-R^{(s)})\Psi^{*}_{class}(\{x_{i}\},\{X_{j}\},t',t)]$ but its internal dynamics  will record differently when the event took place.\footnote{The notation for $\Psi^{*}_{class}$ now includes both the current time, $t$, and the time of the adsorption, $t'$, since this time of internal dynamic response to adsorption is now variable.}  This makes one worry that we will have strong superposition among these different slices.  Even a very small difference in the final \textit{internal} configuration (from acoustic recoil and heating) with time with respect to of a single coordinate label is enough to destroy long term overlap if the localized peak width is narrower than it.   For now we consider this to be a constraint on the initial data.  

This leads us to conjecture the solution is
\begin{equation}
    \Psi(x,\{x_{i}\},\{X_{j}\},t)=
    \begin{cases}
      \hat{\mathcal{S}}\psi(x,t)\cdot \Psi_{class}(x,\{x_{i}\},\{X_{j}\},t) & \ t<t_{1} \\
      \hat{\mathcal{S}} \psi(x<0,t)\cdot \Psi_{class}(x,\{x_{i}\},\{X_{j}\},t)+\\
      \hat{\mathcal{S}}\int_{t_{0}}^{t}  dt' [j_{\perp}(R^{(s)},t')e^{i\phi(R^{(s)},t')}b(x-R^{(s)})\Psi^{*}_{class}(\{x_{i}\},\{X_{j}\},t';t)] & t_{1}<t<t_{2}\\
      \hat{\mathcal{S}} \int_{t_{0}}^{t1} dt' [j_{\perp}(R^{(s)},t')e^{i\phi(R^{(s)},t')}b(x-R^{(s)})\Psi^{*}_{class}(\{x_{i}\},\{X_{j}\},t';t)] & t_{2}<t
    \end{cases}
  \end{equation}
where $j_{\perp}(x,t)$ is the flux of $\psi(x,t)$ inwards and perpendicular to the surface 
and $\phi(x,t)$ is its phase. (This flux is $j=-\frac{\hbar}{m}\Im\psi^{*}\nabla\psi$ if the flux is normal to the surface \textit{and} the surface is at rest.)  
 This ensures proper normalization of the function over space.  The function $b(x)$ is a normalized bump function describing the localization of the adsorbed particle.  
This solution gives an intermediate product state with the remaining free part of the wavefunction and the screen with the previous adsorbed history.\footnote{We should note that the spatial granularity scale is not sufficient to determine the slicing granularity of the function here.  There are two potentially relevant time scales: the reaction time and the sound crossing time across an atomic separation.  The latter is the one we conjecture is the relevant one here.}    

Now let us consider a more general $\psi(x,t)$.  The previous limits on its oscillation rate still apply but now we allow it to be more generally distributed in space.  The solid body is now allowed to have an irregular curved surface $S$ and the body is allowed to move freely and deform so long as the density of adsorbing sites on its surface does not change.  Let the surface sites be denoted by the points $R^{(s)}(t)$ with outwards surface normals $n^{(s)}(t)$.  To compute the sliced classical solid wavefunction we track the incident flux and phase of $\psi$ to the surface within the granularity sweep of each site.  Essentially, the surface captures all the amplitude it sweeps across and at each such event the system is sliced into a noninterfering history.  We must now include the possibility that the unadsorbed parts of $\psi$ can be altered in its evolution by the previous motions of the body and its adsorption history.  This is a significant complication.  Furthermore, the wavefunction incident on our measurement device may be moving rather slowly so that our device can move relatively quickly and so alter the interference of the transmitted portions that other portions of the device can measure later.  A specific example is shown in Fig.\ref{fig:device}.

\begin{figure}
\bc\includegraphics[width=10cm]{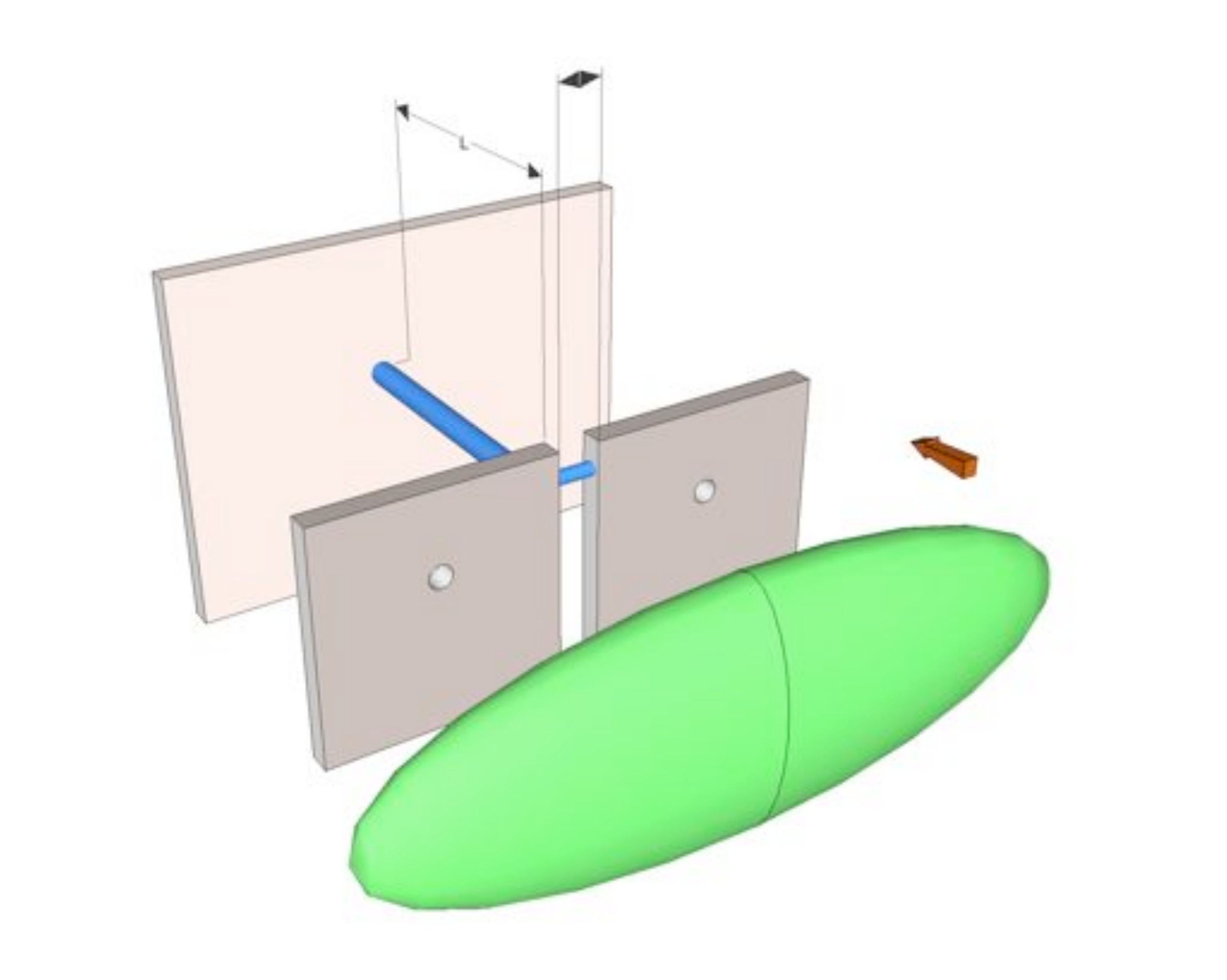}\ec
\caption{\label{fig:device} A measurement device which has holes to allow partial transmission and interference.  The separation of the holes and distance to the farther measuring portions of the device can change in time faster than the characteristic speeds of the wavefunction itself.} 
\end{figure}

Our notation is getting rather cumbersome.  Let us now adopt the following shorthand for our macroscopic body. $\Psi_{class}(R^{(s)}(t),n^{(s)}(t);t',t)$ gives the evolving block under the external and internal motions.  The surface sites and normals are retained but all other coordinates are suppressed.  
To describe the total system at any time we are looking for solutions that are a sum of noninterfering adsorbed states with different histories that are then in a product with the remaining free part of $\psi$.  Such a state looks like a sum of terms like 
\begin{align}
\Psi_{pre}=\psi(x,t) \Psi_{class}(R^{(s)}(t),n^{(s)}(t);t)
\end{align}
and
\begin{align}
\Psi_{post}(s',t')=\rho(R^{(s')},t')e^{i\phi(R^{(s')},t')}b(x-R^{(s')}(t))\Psi^{*}_{class}(R^{(s)}(t),n^{(s)}(t);t,t')
\end{align}
where the $\rho$ and $\phi$ factors are a function of the value of $\psi(x,t)$ at the location and instant where contact with the solid occured.  The index $s'$ is the site where adsorption has occurred and $s$ is a generic site index.  The notation $\Psi_{pre}$ indicates a preadsoprtion history and $\Psi_{post}(s',t')$ indicates a post-adsorption history that occured at site $R^{(s')}$ at time $t'$.  The $\tilde{X}$, $x$ and $t$ parameters are suppressed.  

These results are complicated by the fact that the moving measurement device can cast a ``shadow'' across the incident $\psi$ and introduce new gradients at regions of strongly changing support due to the binding of amplitude to the body.  Energy, momentum and the like must all be conserved in this process so these gradients must be paid for from an enhanced binding energy where such tangential or rapidly oscillating motions of $\psi$ and surface motion occur.  For a specific example, the surface could have a hole through which $\psi$ can pass.  It can then pick up lateral motions just as a light beam will but the strong curvature induced near the wall must come from the binding energy at the edges.  This means that we cannot simply evolve the surface sites through the path of $\psi$ and ignore changes in $\psi$ that may occur as a result since these may affect later measurements for other slices.  A distressing consequence is that we cannot evolve $\Psi_{pre}$ independently of the $\Psi_{post}$ fields where events have already occurred even though $\Psi_{pre}$ remains in the form of a well defined symmetrized product.

The separation of the wavefunction into long separated slices suggests we update Proposition A to allow for such harmless superpositions.

{\bf Proposition B}: The classical state of (condensed) matter can contain slicing over multiple delocalized coordinates as long as the delocalization rate of each ``classical component'' is too slow and the gradients between them too slight to give significant currents over observable time scales.  

This allows us to consider classical matter with wavefunctions of the form
\begin{align}
\Psi=\sum_{i}^{N}\Psi_{class}^{i}
\end{align}
where each $\Psi_{class}^{i}$ is a classical component with well defined shape, center of mass localization, translation, rotation and sufficiently distinct internal structure so that no significant overlap occurs among the other componenets for long times. In the case of nonequilibrium solids and liquids, we will allow different internal flow fields and temperature gradients.  These must be all independently evolving for long times.  

Using this definition of classical matter and its dictionary-like translation to a quantum description we now suggest a completed measurement hypothesis.  

{\bf Measurement Hypothesis}:  At the point and time of contact of a delocalized particle with a classical component of condensed matter, the body becomes sliced into a set of independently evolving components indexed by the phase of the wavefunction at the moment of contact.  This persists for long times due to the large mass of the bound system.  The local energy, momentum and other conserved quantities of the particle are locally deposited in the sliced body at that time and evolve in an effectively classical fashion from then on.  The remaining parts of the wavefunction exist in superposition with the unaltered body as a product function.  The slicing may introduce local curvature in the unadsorbed part of the wavefunction and the price for this comes from the local binding energy.  Energy, momentum, angular momentum, spin, and center of mass motion are not conserved in a given slice but norm, mass and charge are. 

This hypothesis is built on the property of classical matter that allows all measurements to be derived from local processes in space and time.  This is independent of the awareness of any sentient observer.  This might be surprising given the variety of measurement operations we generally think about.  In the case of a helicity separated photon or a Stern-Gerlach spin separated charge particle we see that the action of measurement is a local space time action.  Calorimeters measure the energy deposited from such an action.  Velocity measurements are often taken from wire array data which gives a transversely localized particle path with some curvature.  This seems to be a sequence of localization events that condense slowly compared to the rate the particle passes.  Cloud chamber experiments always involve a surrounding solid material that the particle passes in or out of.\footnote{Tracks in cloud chambers have been knows to exhibit lateral spreading consistent with delocalization.  This suggests that either before or after the particle entered the chamber, a slicing event occurred.  We will later see that gases are intrinsically delocalized objects that do not produce a granularity scale or natural slow partitioning for measurement. }    Note that this measurement hypothesis is testable both experimentally \textit{and} theoretically by direct investigation of the Schr\"{o}dinger equation.  This sets it apart from other approaches with the possible exception of decoherence which itself has yet to resolve basic questions.  

This hypothesis seems a little nonspecific.  We ultimately need an extension of the Schr\"{o}dinger equation of motion for $\psi$ and classical ones for the macroscopic body (which may include preprogrammed changes in response to adsorption events or actions of a sentient being) which then well describe the motion of the net system from the true many body Schr\"{o}dinger equation.  We make explicit the hypothesis with the following rules.
\begin{enumerate}
\item $\Psi_{pre}$ evolves according the Schr\"{o}dinger equation for $\psi$ and Newton's laws for the body subject to sink terms at the surface of the body due to normal currents there.  This ensures no $\psi$ amplitude exists in the interior of the macroscopic body.  (The evolution \textit{at} the surface is to use derivatives in the external local half-space so no discontinuities exist.)
\item $\Psi_{post}$ states evolve in a Newtonian fashion with a fixed relative phase amplitude and weight defined by the $j_{\perp}$ flux and phase at the surface at each time $t'$ in the past.  
\end{enumerate}
In the case that the binding energy density of adsorption is much greater than the kinetic energy density of the incident $\psi$ this seems like a reasonable set of assumptions since the energy conservation constraint does not need to force a local change in free $\psi$ evolution from neighboring collapse events.  

Let us consider the implications of our measurement theory.  It certainly does give the original Born interpretation result in the case of a steady beam with a static flat screen for a measurement device when evolution does not continue to the extent that our slices start to overlap.  No operator for the general position measurement problem seems to exist.  The evolution of the incident particle wavefunction will not always be independent of the history of the device that measures it.  

We can now compare what we might expect from a ``collapse of the wavefunction'' point of view.  The rapid transfer of mass over large distances has not only causality but energy problems.  Presumably, this would have to come from somewhere.  The net shift in the center of mass motion requires a back reaction on the part of the measurement device.  None of these effects have been detected.  Most importantly, the collapse interpretation requires an operator exist to describe the result of measurement from the collapse.  As we noted above, the history dependence of the device should affect the ``uncollapsed'' portion of the incident wave and there can be energy imparted to this wavefunction from the binding energy to it as well.  These give distinct features that can distinguish these models experimentally.  It seems odd that purely Schr\"{o}dinger evolution of a many particle system should reproduce the standard interpretation in typical cases but cause some more dynamic ones to be so different.

\subsection{Other Examples}
\subsubsection{Accelerating Surfaces}
The temporal aspect of measurement is not clear in the usual quantum formalism.  We can elucidate it with a simple 1D case.  Consider an advancing broad packet, corresponding to a single particle, of amplitude $A$ of length $L$ and traveling with velocity $v_{g}$ in the x-direction.  Let a measurement device, in this case a large flat surface perpendicular to x be 
at rest at the origin and at time $t_{1}$ undergo a large impulse that moves it rightwards at velocity $v\gg v_{g}$.  The probability that the particle is adsorbed is the integral of the amplitude squared over the time the wave and the surface intersect.  If a measurement occurs it is recorded at a time $t<t_{1}$ and any further measurement of it later at some point $x_{1}>0$ by another device will be at a time that is $t'\ge x_{1}/v_{g}$.   

\subsubsection{Velocity Measurement}
Consider the motion a packet normally incident on two parallel thin material plates that leave detectable heat or structural changes as a particle transiently binds then continues on.   In this case, the particle is localized but the momentum and energy absorption by the system is considered to be small compared to its net energy.  The transverse motion is constrained by the event and we have a sliced set of evolving beams after the first surface.  The transverse localization in each slice costs energy and this must come from somewhere and the only candidates are from the longitudinal kinetic motion and the internal energy of the plates.  Furthermore, the event leaves a time signature so also localizes longitudinally.  This sets bounds on the minimal $\Delta v$ of the packet and possibly the width of the transverse localization.  Assuming this is still small, the packets will evolve from each site at a well defined $t$ and a spread in $v$.  These spread to reach the second plate with a corresponding spread in arrival time according to its time in transit.  

The final many body wavefunction is a sliced set from the first place and then a further slicing of each from interaction with the second one.  Note that the temporal granularity scale $\Delta t=d/v_{s}$, where $d$ is the atomic separation and $v_{s}$ is the velocity of sound, give a minimum bound on the spread in velocity independent of $\hbar$.

\subsection{Uncertainty Relations}
The role of the uncertainly relations has been argued with no great resolution though they have practical value in placing bounds on experimental results or allowed initial data.  In a deterministic sense, $\sigma_{x}\sigma_{p}$ constraints are nothing more than a statement relating packet width to frequency.  Landau has been quoted as saying ``There is evidently no such limitation-I can measure the energy, and look at my watch; then I know both the energy and the time!'' \cite{Peierls} Bohr and Rosenfeld \cite{Bohr} have given a long discussion on the meaning $\sigma_{E}\sigma_{t}$ constraints.  In a theory that assumes the system is a wave at all times, the relevance of such relations is not clear.  It might be more productive to look at details surrounding the measurement process (since this model gives the specifics to do so) and investigate the detailed reshaping of packets and associated transfer of conserved quantities between measurement device and field.  This will no doubt give many inequalities tailored to the particular configuration at hand.  

As far as the energy of the energy-time relation, the natural place to look for meaning is in the Fourier transform of the time coordinate since this is how the position-momentum relation is derived.  Localizing the temporal oscillations of a packet in time with fixed x is problematic because it violates energy conservation.  Such a localization at a point requires lateral energy flux in the spatial directions.  If we localize a particle in space without a potential we have a similar problem with momentum flux.  To get a stationary analog, we need to turn up a potential uniformly over all space that drives down the oscillations of the packet at $t=\pm\infty$.  The extent to which we get a set of localization in time of our oscillatory region is bounded by the uncertainty in the frequency by $\sigma_{t}\sigma_{\omega}\ge\frac{1}{2}$.  This frequency can be related to the \textit{kinetic} energy as $\sigma_{t}\sigma_{{KE}}\ge\frac{\hbar}{2}$.  We should note that this is only possible for massive particles and the oscillation frequency $\omega$ we assign to it in the Schr\"{o}dinger equation has already hidden the $mc^{2}$ contribution present in the Dirac solutions.  

\subsection{Decoherence}
Decoherence is a proposed many body wavefunction process of creating the appearance of collapse\cite{Schlosshauer}.  The off-diagonal elements of the density matrix are assumed to vanish under natural evolution.  It has been objected by John Bell on the grounds that resolve problems ``for all practical purposes'' \cite{Jammer} but leave oscillations present that could be uncovered through an expensive and time consuming measurement.  
The notion of eigenstate is still explicit in its formulation so seems quite far removed from our picture here of particular class of wavefunctions corresponding to classical objects.  The ``collapse time'' a measure of the time for the off-diagonal components of the density matrix to vanish.  In our described measurement process such a notion clearly has no meaning as will be elaborated below.  Another way to see that these two pictures are not equivalent is in the observation that the quantum probabilities don't always follow the usual projector description as noted in the examples.

An essential difference between the method discussed in this paper and decoherence is the role of the eigenstate and that classical matter only has the right kinematic features when it is very far from the eigenstates of the actual Hamiltonians.  Interestingly, the kind of specialized Hamiltonians we construct in condensed matter physics for particular crystal structures, in some sense, incorporates this as they implicitly assume well defined atomic locations.  It has been overlooked that the solutions of these Hamiltonians are different from the true ones in profound ways.  
In understanding thermal equilibrium and transport, the true eigenstate basis will turn out to be similarly irrelevant but also point to a new way to interpret quantum statistical ensembles without the usual contradictions and ad hoc nature.  

\subsection{Origins of the Initial Data}
So far we have no idea how the initial data suggested by Proposition B arises.  We can consider an ``Ehrenfest-gas'' of localized clusters at most experimental densities.  This moves briefly in a classical fashion but rapidly delocalizes over distances far greater than the interparticle separation.  Furthermore, this process is independent of the thermal energy of the initial system.  For this reason, the classical description of thermodynamics and hydrodynamics we arrive at from a kinetic theory of bouncing billiard balls seems completely irrelevant.  That the numerical results we obtain do an excellent job of modeling most aspects of high temperature gases, seems almost like a red herring or, at best, a hint towards some similarity between the dynamics of a ``typical'' delocalized many body wavefunction at high enough energy and the classical problem.  

We are not going to address thermodynamic equilibrium here in depth from this point of view, but to comment that such a condition is problematic for a genuine quantum system where the particular superpositions of eigenstates for an isolated system are preserved for all time.  However, this sort of situation is exactly the initial data from which our sun and planets, hence all available condensed matter presumably arose.  

In a strongly delocalized gas, condensation involves ejecting energy in the form of radiation and ejected particles to allow the forming matter to attain a lower energy state.  The discussion presented here does not include radiation, since the particle number is assumed fixed so we are limited in this respect.  In the case of a forming dust particle, we see that a forming pair has no well defined orientation.  Further adsorption leads to a superposition of many shapes and binding site choices.  
As these clusters grow, rebounding particles that leave the system provide more low mass parameters to slice the systems and keep the growing macroscopic superpositions increasingly isolated.  This provide a qualitative picture of the origins of the state but clearly leave much room for further analysis.  Interestingly, nucleation theory \cite{Wolk} is still in very unsatisfactory state giving another reason why such a more specific picture of quantum and classical transitions is germane.

\subsection{Further Complications}
Schr\"{o}dinger cat type arguments involve the alteration of classical behavior in response to a quantum measurement.  We can certainly do this with our moving measurement device above.  The important distinction is that the shadow effects on the incident $\psi$ depend only on the history of the device up until a selection event is made.  Everything after that has no effect on the later adsorption events.  

The particle adsorption does not really create the slicing of the system.  The measurement device is already sliced into copies indexed the the values of the incident wavefunction $\psi(x)$.  The adsorption event simply causes the identical copies to exhibit different behaviors.  Decoherence approaches are concerned with the time for the collapse to occur.  For our description, no such notion arises.  There are however other timescales.  For example, the time for the binding to occur, typically measurable by the length of radiation losses.  Also the local elastic changes that radiate out from the site of adsorption and equilibrate at the speed of sound over a number of crossing times of the body.  For relativistic measurement devices this is unchanged.  However, it is presumed that there is some equilibration over the totality of the measurement device before the events occur.  In the case of a gas there is no collective binding that generates a set of localized sites for this to occur.  For a condensing solid or a flowing liquid, the partitioning of the space into a well defined set of slices that would be consistent with Proposition B may not have arised.  In laboratory work this may not generally arise, except for the case of strongly isolated mesoscopic solids that might arise in future trapped gas and diamagnetic levitation work, but it may be a concern for astrophysical systems and nucleation.  

Macroscopic superpositions such as ``Schr\"{o}dinger's cat'' produce no paradoxical effect because we have been very specific in the sort of wavefunctions to correspond to the system.  We can see that the energy changes with and without the superposition are very small.  The measurment induces long lasting partitions of observer histories with the observed probabilities for static devices.  If we consider more general and naive macroscopic superpositions such favorable conditions do not hold.  Consider a block with core locations at $\tilde{R}$ and a slightly shifted or rotated lattice $\tilde{R}'$.  If we now make a naive superposition of the cores as $\Psi_{class}+\alpha\Psi_{class}'$ we have some cores that will in general be much closer than the interatomic separation and others sitting between interstitial sites.  When we construct the corresponding electron orbitals that form the bonds to hold the lattice together the energy per bond is increased by an amount on the order of the deepest bound electron energy per electron.  Such a solid is certainly not stable and the energy contributions per atom are much greater than orbital binding energy.   

The superselection properties of quantum mechanics have never had a clear resolution.  Symmetry breaking depends on it but why nature chooses one option is never clear.  Consider the case of a single domain magnetic grain isolated in space to which we had or subtract heat.  When we cool below the Curie temperature it must select a direction to magnetize.  The superposition of both up and down spin solutions is permissible based on Schr\"{o}dinger evolution yet we don't observe these (though ferromagnetic transitions in ultracold gases may be a place to look).  When the magnetic field of the body interacts with the external world and its own atoms it polarizes spins and alters electron orbits through paramagnetic and diamagnetic effects.  It seems that the property of $\Psi_{class}$ to slice into functions that evolve independent of each other forces observers to observe living in a world where only one of the two options occurs.  Reheating above $T_{c}$ seems like it would lead to interference and a ``revival'' of interaction among the alternative slices to interfere and create unexplainable changes or forces to the observers and even a conflict in the memory of a recording device.  The magnitude of internal structural changes in such a system seems crucial in determining if this occurs.  Phase revival has been seen in bose gases\cite{Dolfovo}.  If we can condense such gases into solids while preserving their isolation and delocalization, this might be an ideal testbed to study the process of superselection.  The presence of superselection is more ubiquitous than often noted.  The famed ``Mexican hat'' potential associated with symmetry breaking does not fix a given value.  Specific systems of importance that exhibit superselection are the quantum blockade, the selection of a particular supercurrent in superconductors, superfluid vortices in liquid He, and the quantum Hall effect. 

The metastable nature of the classical states we describe introduce a natural arrow of time to the system.  Long before we encounter Poincare cycle like behavior we should confront the long time limits that compromise the classical nature of the solids and observers that comprise it.  This can occur just but allowing time to evolve but iterative measurements will enhance it.  
The slicing of our many body wavefunction cannot go on indefinitely.  Eventually delocalization causes these to interfere.  Delocalization of these classical states undermines the very notion of a body with well defined shape, orientation and even particle number.  In the theory of quantum information, information itself is recognized as a physical quantity.  General wavefunctions lose the ability to durably store information in a binary fashion and thus to manipulate it in a repeatable manner.  The long time limit seems to have problems bigger than the Poincare cycle and the arrow of time.  It loses these essential properties on which life and consciousness themselves seem to depend. 

\section{Conclusions}

Through the specificity of constructed wavefunctions and a detailed analysis of which ones correspond to a classical description of the observed macroscopic world, we have presented a notion of measurement that subsumes the usual Born statistics and gives a more general description of measurement.  In the course of this, we have given a specific picture of phonon excitations in terms of the many body wavefunction of a solid and discussed the fundamental differences between the fluid states of matter and solids, proposing that metastable condensed matter states provide the foundation both the classical appearance of the world and a foundation for information and knowledge in general.  

The morals of this approach are that all ``measurements,'' specifically the interaction of small delocalized objects with large classical seeming ones, are essentially space and time measurements.  Other conserved quantities are locally deposited along with this event so that the conservation of them in the many body function arises from the sum over all slices.  
This ``intrinsic'' theory of measurement implies that the incident particle part of the wavefunction undergoes rather gentle changes in the act corresponding to localization from the granularity scale to the size of the bound peak (i.e.\ the width of $\psi_{0}$ in eqn.\ \ref{class}) at each site and the apparent abrupt changes are due to the bifurcations of the observed history as indexed by the incident wavefunction itself.  The massive difference in sizes of quantum and classical phase spaces make this possible due to a presumption that condensed matter wavefunctions tend to distribute themselves rather sparsely in this space due to interaction during formation with large quantities of ejected gas. 

This picture begs at least as many questions as it answers.  It suggests fundamentally new sorts of measurements corresponding to richly dynamic measurement devices compared to the motions of the incident wavefunction.  Generally solids and gases are considered on a firm quantum foundation with liquids being more questionable.  Here we have shown that the hydrodynamic behavior and kinematic constraints of gases presumed to be in the ``classical regime,'' despite tending to rapidly evolve strong quantum delocalization, will require a more serious reconsideration.  Statistical mechanics and these hydrodynamic concerns will be discussed in a following paper.  Given the importance placed on the thermodynamic and hydrodynamic approaches to ultracold gases, a more clear understanding of these aspects of the high energy behavior of gases seems urgent.  

The author gratefully acknowledges conversations with Thomas Sch\"{a}fer and Lubos Mitas.

\bibliographystyle{plain} 
\cleardoublepage
\normalbaselines 
\clearpage
\bibliography{References} 
\endgroup

\end{document}